\pdfoutput=1																					

\documentclass[3p,times,twocolumn]{elsarticle}

\newcommand{\no}{\noindent}
\newcommand{\myeq}[3]{\vspace{#2} \begin{equation} \hspace{#1} #3 \end{equation} \vspace{0cm}}
\newcommand{\vsp}[1]{\vspace*{#1}}
\newcommand{\hsp}[1]{\hspace*{#1}}
\newcommand{\Op}{\mathcal{O}}

\usepackage{amssymb}

\usepackage[figuresright]{rotating}

\begin{document}

\begin{frontmatter}




\title{Constraints on 3-flavor QCD order parameters from $\eta\to\pi^+\pi^-\pi^0$ decay width\tnoteref{label1}}
\tnotetext[label1]{This work was supported by project MSM0021620859 of the Ministry of Education of Czech Republic.}


\author{Mari\'{a}n Koles\'{a}r, Ji\v{r}\'i Novotn\'y}

\address{Institute of Particle and Nuclear Physics, Faculty of Mathematics and Physics, Charles University in Prague, CZ-18000 Prague, Czech republic}

\begin{abstract}

Quark condensate and pseudoscalar decay constant in the chiral limit are the principal order parameters of spontaneous chiral 
symmetry breaking in QCD. Yet their three flavor values are still only weakly constrained by analyses using experimental  
data. We try to obtain such constraints by statistical methods from the decay 
width of the $\eta$$\,\to\,$$\pi^+\pi^-\pi^0$ decay in the framework of resummed chiral perturbation theory. We rely on recent  		
estimates of the isospin violating parameter $R$, which is proportional 
to the difference of the u and d quark masses. Alternatively, we relax the assumption and try to extract information on $R$ as well.

\end{abstract}


\end{frontmatter}



Spontaneous breaking of chiral symmetry (SB$\chi$S) is a prominent feature of the QCD vacuum and thus its character has been under discussion for a long time \cite{Fuchs:1991cq,DescotesGenon:1999uh}. The principal order parameters are the quark condensate and the pseudoscalar decay constant in the chiral limit

\myeq{0cm}{0cm}{
	 \Sigma(N_f) = \langle\,0\,|\,\bar{q}q \,|\,0\,\rangle\,|_{m_q\to 0}\,,}
\myeq{0cm}{-0.25cm}{
	 F(N_f) = F_P^a\,|_{m_q\to 0}\,,\quad   i p_{\mu}\, F_P^a\ =\ \langle\,0\,|\,A_{\mu}^a\,|\,P\,\rangle,}

\no where $N_f$ is the number of quark flavors $q$ considered light, $m_q$ collectively denotes their masses. $A_{\mu}^a$ are the QCD axial vector currents, $F_P^a$ the decay constants of the light pseudoscalar mesons $P$.

Chiral perturbation theory ($\chi$PT) \cite{Weinberg:1978kz,Gasser:1983yg,Gasser:1984gg} is constructed as a general low energy parameterization of QCD based on its symmetries and the discussed order parameters appear at the lowest order of the chiral expansion as low energy constants (LECs). Interactions of the light pseudoscalar meson octet, the pseudo-Goldstone bosons of the broken symmetry, directly depend on the pattern of SB$\chi$S and thus can provide information about the values of these observables.

A convenient reparameterization of the order parameters, relating them to physical quantities connected with pion two point Green functions, can be introduced \cite{DescotesGenon:1999uh}

\myeq{0.25cm}{0cm}{
	Z(N_f) = \frac{F(N_f)^2}{F_{\pi}^2},\quad
	X(N_f) = \frac{2\hat{m}\,\Sigma(N_f)}{F_{\pi}^2M_{\pi}^2},}

\no where $\hat{m}\,$=$\,(m_u\,$+$\,m_d)/2$. $X(N_f)$ and $Z(N_f)$ are limited to the range (0,\,1), $Z(N_f)$\,=\,0 would correspond to a restoration of chiral symmetry and $X(N_f)$\,=\,0 to a case with vanishing chiral condensate. Standard approach to chiral perturbation series tacitly assumes values of $X(N_f)$ and $Z(N_f)$ not much smaller than one, which means that the leading order terms should dominate the expansion.

The two flavor values were determined some time ago \cite{DescotesGenon:2001tn}

\myeq{0cm}{0cm}{X(2)=0.81\pm0.07,\quad Z(2)=0.89\pm0.03,}

\no while the three flavor ones are constrained from above by the so-called paramagnetic inequalities \cite{DescotesGenon:1999uh}

\myeq{0.5cm}{0cm}{\label{param}X\equiv X(3)<X(2),\quad Z\equiv Z(3)<Z(2).}

\no NNLO standard chiral perturbation theory fits have obtained conflicting results - the classic ``fit 10" values are $Z$\,=\,0.89 and $X$\,=\,0.66 \cite{Amoros:2001cp}, while the recent update \cite{Bijnens:2011tb} provides $Z$\,=\,0.50, $X$\,=\,0.51 for the main fit and $Z$\,=\,0.44, $X$\,=\,0.29 for the linear $K_{e4}$ fit. Resummed $\chi$PT analyses using $\pi\pi$ \cite{DescotesGenon:2003cg} and $\pi\pi$+$\pi K$ \cite{DescotesGenon:2007ta} scattering data only obtained relatively weak constraints, $Z$$\,>\,$0.2, $X$$\,<\,$0.8 at 68\% C.L. in the $\pi\pi$+$\pi K$ case \cite{DescotesGenon:2007ta}. Using the resummed framework to fit lattice QCD data \cite{Bernard:2010ex,Bernard:2012fw,Bernard:2012ci} provides results in the range $Z$$\,\sim\,$0.5-0.7, $X$$\,\sim\,$0.3-0.5, the latest values being $Z$\,=\,0.54$\,\pm\,$0.06, $X$\,=\,0.38$\,\pm\,$0.05 \cite{Bernard:2012ci}. 

The mentioned alternative approach, dubbed resummed $\chi$PT \cite{DescotesGenon:2003cg}, was developed in order to accommodate the possibility of irregular convergence of the chiral expansion. The procedure can be very shortly summarized in the following way:

\vsp{-0.1cm}
\begin{itemize}	
	\item[-]	standard $\chi$PT Lagrangian and power counting \vsp{-0.2cm}
	\item[-]	only expansions related linearly to Green functions of the QCD currents trusted \vsp{-0.2cm}
	\item[-] 	explicitly to NLO, higher orders implicit\\ in remainders\vsp{-0.2cm}
	\item[-]	remainders retained, treated as sources of error \vsp{-0.2cm}
	\item[-]	manipulations in non-perturbative algebraic way \vsp{-0.1cm}
\end{itemize}

\no In practice, the two approaches typically differ by reordering of the chiral expansion for the calculated observables. The hope for resummed $\chi$PT is that by carefully avoiding dangerous manipulations a better converging series can be obtained. The procedure also avoids the hard to control NLO a NNLO LECs by trading them for remainders with known chiral order.

Theoretical efforts to explain the $\eta$$\,\to\,$$3\pi$ 
decays reach far back in time. This is an isospin breaking process,
as three isovectors can constitute an isoscalar state only through
the fully antisymmetric combination $\epsilon_{abc}\pi^a\pi^b\pi^c$, which
together with Bose symmetry and charge conjugation invariance leads to zero
contribution to the amplitude.

When a systematic approach to low energy hadron physics was born in the form of $\chi$PT, 
it was quickly applied to the $\eta$$\,\to\,$$3\pi$ decays
\cite{Gasser:1984pr}. The one loop corrections were found to be very sizable, the result for the decay width of the
charged channel was 160$\pm$50 eV, compared to the current algebra prediction of 66 eV. However, 
already at that time there were hints that the experimental value is still much larger. The current 
PDG value is \cite{PDG:1900zz}

\myeq{1.75cm}{0cm}{\Gamma_\mathrm{exp} = 296 \pm 16 \ \mathrm{eV}.}

\no At last, the two loop $\chi$PT calculation \cite{Bijnens:2007pr} has succeeded to obtain a reasonable result for the decay widths. 

The theory thus might seem to converge really slowly for the decays. However, as will be shown elsewhere \cite{Kolesar:prep}, we argue that the four point Green functions relevant for the $\eta$$\,\to\,$$3\pi$ amplitude (see (\ref{Green_f}) below) do not necessarily have large contributions beyond next-to-leading order and a reasonably small higher order remainder is not in contradiction with huge corrections to the width. The width does not seem to be sensitive to the details of the Dalitz plot distribution, but rather to the value of leading order parameters - the chiral decay constant, the chiral condensate and the difference of $u$ and $d$ quark masses, i.e. the magnitude of isospin breaking. Moreover, access to the values of these quantities is not screened by EM effects, it was shown that the electromagnetic corrections up to NLO are very small \cite{Baur:1995gc, Ditsche:2008cq}. This observable thus seems to be quite suitable for the application of the resummed $\chi$PT methodology and therefore we employ it to try to extract information about the character of the QCD vacuum.

Our calculation closely follows the procedure outlined in \cite{Kolesar:2008jr}. 
We start by expressing the charged decay amplitude in terms of 4-point
Green functions $G_{ijkl}$, obtained from the generating functional of the QCD currents. We compute at first order in isospin breaking, 
in this case the amplitude takes the form

\myeq{-0.7cm}{0cm}{\label{Green_f}
	F_\pi^3F_{\eta}A(s,t,u)
		= G_{+-83}-\varepsilon_{\pi}G_{+-33}+\varepsilon_{\eta}G_{+-88} + \Delta^{(6)}_{G_D},\ }
		
\no where $\Delta^{(6)}_{G_D}$ is the direct higher order remainder to the 4-point Green functions. 
The physical mixing angles to all chiral orders and first in isospin breaking
can be expressed in terms of quadratic mixing terms of the generating functional to NLO
and related indirect remainders

\myeq{-0.5cm}{0cm}{
	\varepsilon_{\pi,\eta} = -\frac{F_{0}^{2}}{F_{\pi^0,\eta}^{2}}
		\frac{(\mathcal{M}_{38}^{(4)}+\Delta_{M_{38}}^{(6)}) - 									
		M_{\eta,\pi^0}^{2}(Z_{38}^{(4)}+\Delta _{Z_{38}}^{(6)})}
		{M_\eta^2-M_{\pi^0}^2}.}

In accord with the method, $\Op(p^2)$ parameters appear inside loops, while
physical quantities in outer legs. Such a strictly derived amplitude has an
incorrect analytical structure due to the leading order masses in loops, cuts and poles being in unphysical positions. To account
for this, we exchange the LO masses in unitarity corrections and chiral logarithms for physical ones, as described in \cite{Kolesar:2008jr}.   

The next step is the treatment of the LECs. As discussed, the leading order
ones, as well as quark masses, are expressed in terms of convenient parameters

\myeq{-0.6cm}{0cm}{
	Z = \frac{F_0^2}{F_{\pi}^2},\ \
	X = \frac{2\hat{m}\,F_0^2B_0}{F_{\pi}^2M_{\pi}^2},\ \
	r = \frac{m_s}{\hat{m}},\ \
	R = \frac{m_s-\hat{m}}{m_d-m_u}.\ \ }
	
\no At next-to-leading order, the LECs $L_4$-$L_8$ are algebraically reparametrized in terms
of pseudoscalar masses, decay constants and the free parameters $X$, $Z$ and $r$ using chiral expansions of 
two point Green functions, similarly to \cite{DescotesGenon:2003cg}. Because expansions are formally not truncated, 
each generates an unknown higher order remainder.

We don't have a similar procedure ready for $L_1$-$L_3$ at this point, therefore we collect
a set of standard $\chi$PT fits \cite{Amoros:2001cp,Bijnens:2011tb,Bijnens:1994ie} and by taking their mean and spread, while 
ignoring the much smaller reported error bars, we obtain an estimate of their influence

\myeq{0.75cm}{0cm}{L_1^r(M_\rho) = (0.60\pm$$0.28) \cdot 10^{-3}}
\myeq{0.75cm}{-0.5cm}{L_2^r(M_\rho) = (0.88\pm$$0.34) \cdot 10^{-3}}
\myeq{0.75cm}{-0.5cm}{L_3^r(M_\rho) = (-2.97\pm$$0.47) \cdot 10^{-3}}

\no As will be shown in \cite{Kolesar:prep}, the results depend on these constants only very weakly.

The $O(p^6)$ and higher order LECs, notorious for their abundance, are implicit in
a relatively smaller number of higher order remainders. We have eight indirect remainders
- three generated by the expansions of the pseudoscalar masses, three by the decay constants and two by the mixing angles.
We expand the direct remainder to the 4-point Green functions around the center of the Dalitz plot $s_0=1/3(M_\eta^2$+$2M_{\pi^+}^2$+$M_{\pi^0}^2)$
		
\myeq{-0.5cm}{0cm}{
	\Delta^{(6)}_{G_D} = \Delta_A+\Delta_B(s-s_0)\,+}
\myeq{0.5cm}{-0.5cm}{+\,\Delta_C(s-s_0)^2+\Delta_D [(t-s_0)^2+(u-s_0)^2]}
	
\no and thus get four derived direct remainders, two NLO and two NNLO ones. As the experimental curvature of the Dalitz plot is very small \cite{KLOE:2008ht}, we argue that for the purpose of integrating over the kinematic phase space in the decay width calculation, the expansion to second order in the Mandelstam variables is sufficient.

For the statistical analysis, we use an approach based on Bayes' theorem \cite{DescotesGenon:2003cg}

\myeq{0.5cm}{0cm}{
		P(X_i|\Gamma_\mathrm{exp}) = \frac{P(\Gamma_\mathrm{exp}|X_i)P(X_i)}{\int \mathrm{d}X_i\,P(\Gamma_\mathrm{exp}|X_i)P(X_i)}\,,}

\no where $P(X_i|\Gamma_\mathrm{exp})$ is the probability density of the parameters and remainders, denoted as $X_i$, having a specific value given the observed experimental width $\Gamma_\mathrm{exp}$. $P(\Gamma_\mathrm{exp}|X_i)$ is the known probability density of observing $\Gamma_\mathrm{exp}$ in an experiment under the assumption that the true values of $X_i$ are known

\myeq{-0.25cm}{0cm}{\label{Bayes}
		P(\Gamma_\mathrm{exp}|X_i) = \frac{1}{\sigma_\mathrm{exp}\sqrt{2\pi}}\,
		\mathrm{exp}\left[-\frac{(\Gamma_\mathrm{exp}-\Gamma(X_i))^2}{\sigma_\mathrm{exp}^2}\right].}
		
\no $P(X_i)$ is the prior probability distribution of $X_i$. We use it to implement the theoretical uncertainties connected with our parameters and remainders. In such a way we keep the theoretical assumptions explicit and under control. It also allows us to test various assumptions and formulate if-then statements.

We assume the strange to light quark ration $r$ to be known and use the lattice QCD average \cite{Colangelo:2010et}

\myeq{2cm}{0cm}{r = 27.8 \pm 1.0.}

As for the remainders, we use an estimate based on general arguments about the convergence 
of the chiral series \cite{DescotesGenon:2003cg}
		
\myeq{1cm}{0cm}{\label{Delta_G}\Delta_G^{(4)}\ \approx\pm 0.3 G,\quad \Delta_G^{(6)}\ \approx\pm 0.1 G,}

\no where $G$ stands for any of our 2-point or 4-point Green functions,
which generate the remainders. We implement (\ref{Delta_G}) by using a normal distribution with $\mu\,$=\,0 and $\sigma\,$=\,0.3$G$ or $\sigma\,$=\,0.1$G$ for the NLO or NNLO remainders, respectively. In contrast with the frequentist approach \cite{DescotesGenon:2007ta}, the remainders are thus not limited by any value, in fact with very high probability at least some of the values exceed 0.3$G$ or 0.1$G$. We have also tested a uniform distribution akin to \cite{DescotesGenon:2007ta}, results are qualitatively similar to the ones with normal distribution, with obtained constraints being stronger. These results will be published later \cite{Kolesar:prep}.

At last, we are left with three free parameters: $X$, $Z$ and $R$. These
control the scenario of spontaneous breaking of chiral symmetry and isospin breaking in our results. In the case of $X$ and $Z$, we only use the constraint from the paramagnetic inequality (\ref{param}) and assume these parameters to be in the range

\myeq{1.25cm}{0cm}{X\in(0,0.9),\quad Z\in(0,0.9).}

\no We use two approaches to deal with $R$. In the first one we assume it to be a known quantity. We use the value

\myeq{2cm}{0cm}{R=37.8\pm3.3,}

\no obtained from a dispersive analysis of $\eta$$\,\to\,$$3\pi$ \cite{Kampf:2011wr}. However, one should be aware that this estimate is based on an assumption that NNLO standard $\chi$PT \cite{Bijnens:2007pr} converges well at a specific kinematic point found in unphysical region. In comparison, \cite{Bijnens:2007pr} arrives at $R$\,=\,42.2. 

Alternatively, we leave $R$ free, or more precisely, assume it to be in a wide range 

\myeq{2.25cm}{0cm}{R\in(20,60).}

We resort to Monte Carlo sampling in order to perform the numerical integration in (\ref{Bayes}). We have used 10000 samples per grid element, the total number of samples being $\sim\,$$10^6$. We have verified the stability with smaller number of samples (1000 per grid element), an in depth stability test is in preparation.

\begin{figure}[t]
	\hsp{0cm}
	\includegraphics[width=0.5\textwidth]{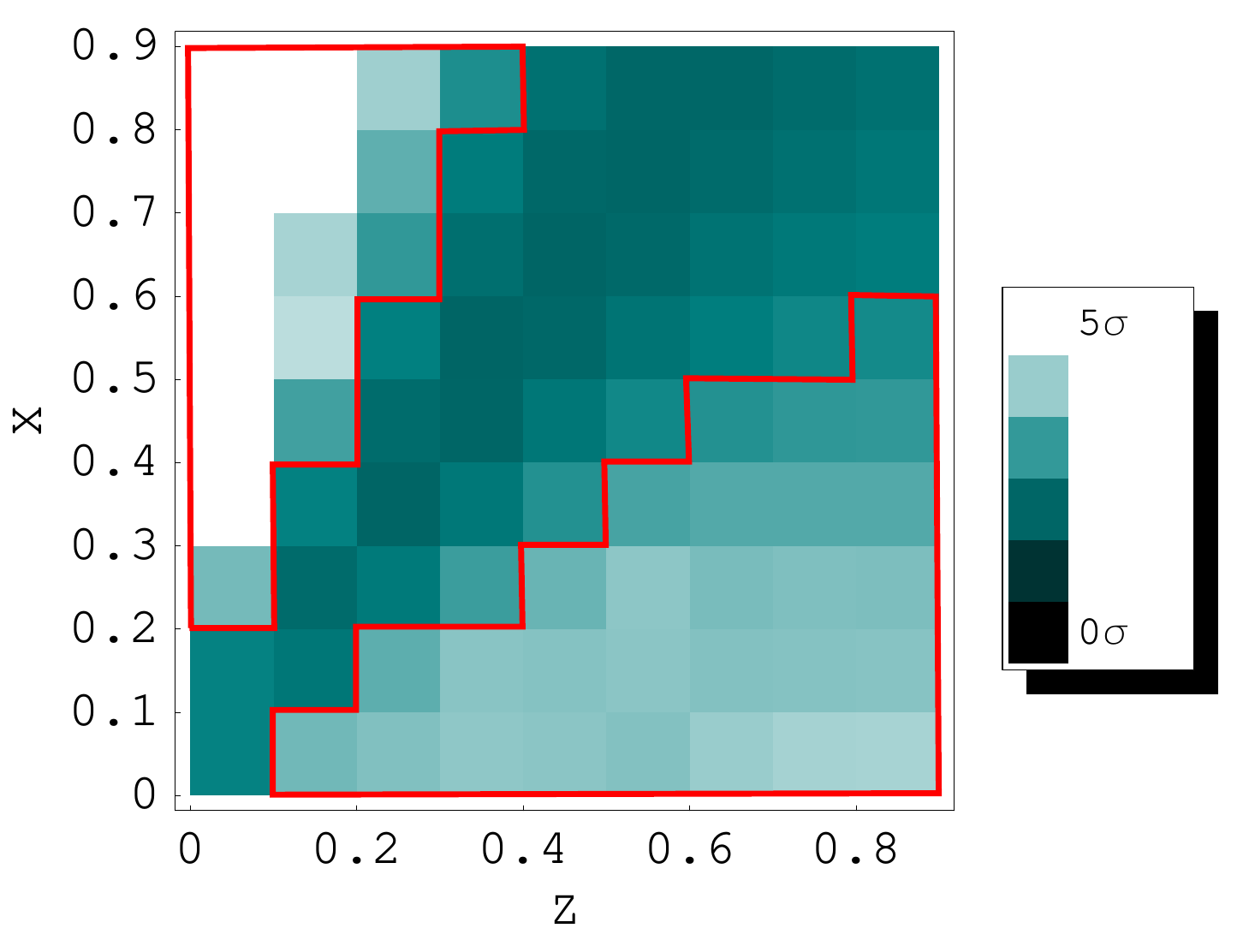}					
	\caption{Probability density $P(X,Z|\Gamma_\mathrm{exp})$ for $R$\,=\,37.8$\,\pm\,$3.3\newline
					 highlighted area: regions with $Y_\mathrm{max}$$\,\leq\,$0.75 and $Y_\mathrm{min}$$\,\geq\,$2$\,$:\newline
					 \hsp{2cm}excluded at $p\,$=96.4\%=2.1$\sigma$ C.L.}
	\label{fig1}
\end{figure}

\begin{figure}[t]
	\hsp{0cm}
	\includegraphics[width=0.5\textwidth]{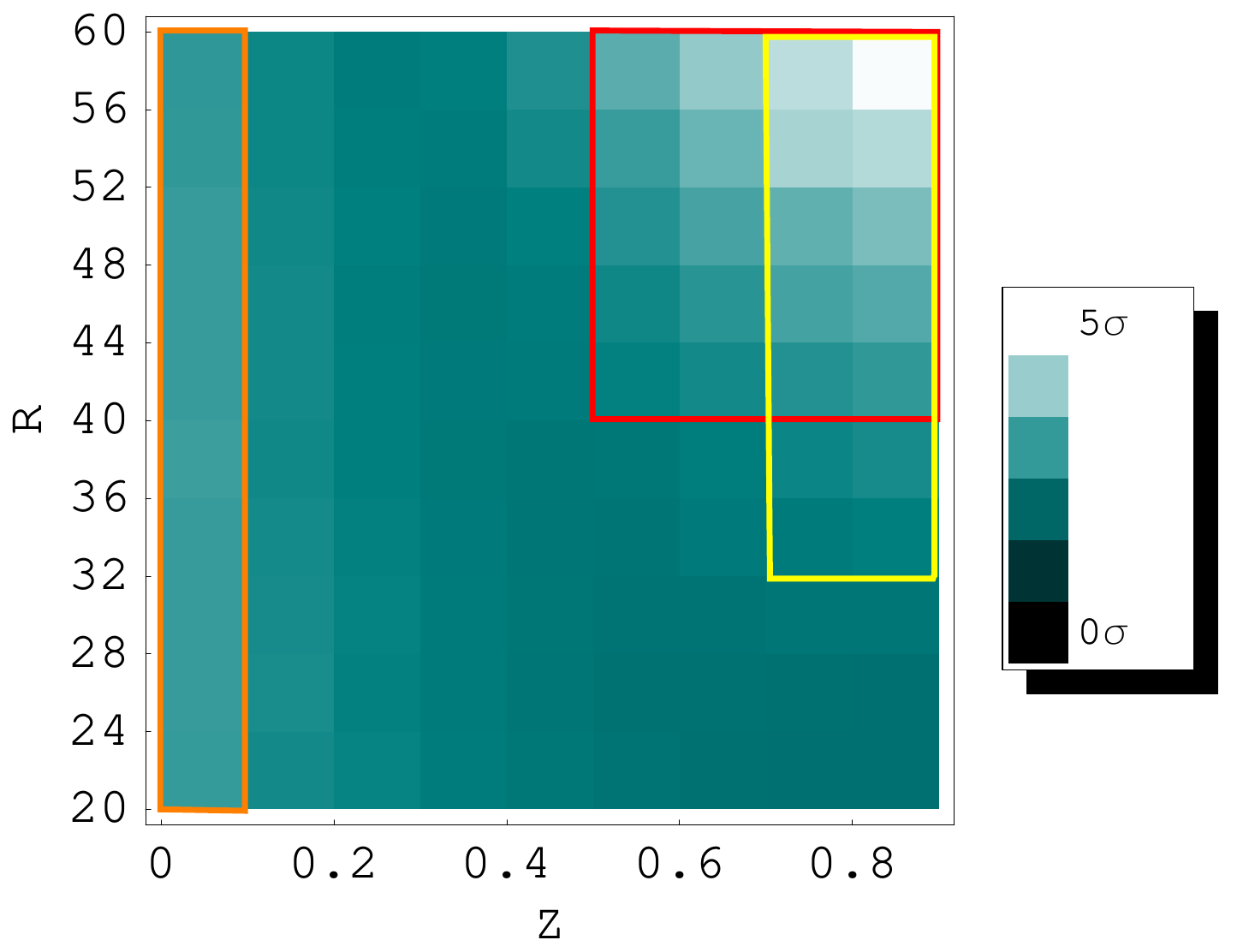}				
	\caption{Probability density $P(R,Z|\Gamma_\mathrm{exp})$ for $X\in(0,0.9)$\newline
					 highlighted areas:\newline
					 \hsp{0.25cm}$Z\,$$>$0.5 and $R\,$$>$40\,: excluded at $p\,$=95.0\%=2.0$\sigma$ C.L.\newline
					 \hsp{0.25cm}$Z\,$$>$0.7 and $R\,$$>$32\,: excluded at $p\,$=94.6\%=1.9$\sigma$ C.L.\newline
					 \hsp{0.25cm}$Z\,$$<$0.1\,: excluded at $p\,$=97.8\%=2.3$\sigma$ C.L.\newline}
	\label{fig2}
\end{figure}

The obtained probability density distributions can be found in figures \ref{fig1} and \ref{fig2}. As can be seen, our first results have shown that the $\eta\to\pi^+\pi^-\pi^0$ decay width is indeed sensitive to $X$ and $Z$. A large 
portion of the parameter space can be excluded at $p$$\,>\,$2.0$\,\sigma$ C.L., given information about $R$. It seems $Y$\,=\,$X/Z$$\,\geq\,$1 is preferred, therefore we have a specific test for $Y$ in preparation. 

As expected, it's hard to constrain $R$ without information on $X$ and $Z$, we have thus obtained conditional constraints.
Assuming $Z$$\,>$0.5 excludes the region $R$$\,>$40 at 2.0$\,\sigma$ C.L. and $Z$$\,>$0.7 excludes $R$$\,>$32 at 
1.9$\,\sigma$ C.L. $Z$$\,<$0.1 can be excluded at 2.3$\,\sigma$ C.L.

As an outlook, we work on an in depth statistical stability test of the Monte Carlo sampling and plan to extend the 
analysis to more parameters and include a wider range of experimental data.

\nocite{*}
\bibliographystyle{elsarticle-num}
\bibliography{Bibliography}







\end{document}